\documentclass[prl,twocolumn,showpacs,floatfix,amsmath,amsfonts]{revtex4}

\newcommand{\cU}{{\cal U}}
\newcommand{\cG}{{\cal G}}
\newcommand{\cE}{{\cal E}}

\newcommand{\cA}{{\cal A}}

\usepackage{epsfig}
\usepackage{graphicx}
\usepackage{amsmath}
\usepackage{amssymb}
\usepackage{dcolumn}
\begin{document}

\title{Long-living Bloch oscillations of matter waves in optical lattices}


\author{M. Salerno$^{1}$}

\author{V. V. Konotop$^{2,3}$}

\author{Yu. V. Bludov$^{3}$}

\affiliation{ $^1$Dipartimento di Fisica "E. R. Caianiello," and
Consorzio Nazionale Interuniversitario per le Scienze Fisiche
della Materia (CNISM), Universit\'a di Salerno, I-84081, Baronissi
(SA), Italy
\\
 $^2$Departamento de F\'{\i}sica, Universidade de
Lisboa, Campo Grande, Edifício C8, Piso 6, Lisboa 1749-016, Portugal
\\
$^3$Centro de F\'{\i}sica Te\'orica e Computacional, Universidade
de Lisboa, Complexo Interdisciplinar, Avenida Professor Gama
Pinto 2, Lisboa 1649-003, Portugal
}

\begin{abstract}
It is shown that by properly designing the spatial dependence of
the nonlinearity it is possible to induce long-living Bloch
oscillations of a localized wavepacket in a periodic potential.
The results are supported both by analytical and numerical
investigations and are interpreted in terms of  matter wave
dynamics displaying dozens of oscillation periods  without any
visible distortion of the wave packet.

\end{abstract}

 \maketitle

The phenomenon of Bloch oscillations (BO), predicted  by
Bloch in 1928 in his celebrated paper on the dynamics of a band
electron in a steady electrical field \cite{Bloch}, represents a  problem of
non exhausted interest. Besides solid state physics, where the
phenomenon has been observed only in recent times after the
development of the superlattice technology~\cite{ExperSolid}, it
is now possible to observe BO also in other fields such as
nonlinear optics, using light beams in arrays of
waveguides~\cite{wavegides} or in photorefractive
crystals~\cite{photo},  and atomic physics, using Bose-Einstein
condensates (BEC) loaded in optical
lattices (OLs)~\cite{ChoiNiu,BEC-bloch,KK}. Apart its fundamental
significance, the interest in BO arises mainly from perspectives
of their practical applications. In this context we mention the use of
BO for metrological tasks, including relatively precise definition
of $h/m$~\cite{metrology} and measurement of forces at the
micrometer scale like the Casimir-Polder force~\cite{sources} and
the gravity~\cite{gravity}. Recently, BO were also suggested as a
tool for controlling light in coupled-resonator optical
waveguides~\cite{Longhi}. In these physical contexts BO have been
observed mainly in the  linear (as first proposed by Bloch) or
quasi linear regimes (where the nonlinearity introduces quantitative but not
 yet qualitative changes). The present experimental settings (both in
nonlinear optics and in BECs), however, allow for attaining essentially
nonlinear regimes quite easily, this actually being a desirable
condition for applications involving localized states.

The fact that BO can exist in presence of interactions
(nonlinearity) was first recognized in the context of nonlinear
discrete systems~\cite{Bloch-discr}. For periodic continuous models of the
nonlinear Schr\"odinger (NLS) type, such as ones describing matter
waves in OLs, the existence of BO becomes more problematic because
of nonlinearity induced instabilities in the underlying linear
system. These instabilities can be simply understood by observing
that in a usual OL at the edges of an allowed band the effective
mass has always opposite signs. This implies that if a Bloch state
is modulationally stable (in presence of a constant nonlinearity)
at one edge of the band, it must be necessarily unstable at the
other band edge~\cite{KS}. Nonlinearity induced instabilities have
been extensively investigated both
theoretically~\cite{KS,stability} and
experimentally~\cite{instab-exper}) and have been identified as
primary cause for the short life-times of BO of matter waves (the
oscillation survives only a few cycles) observed both in
numerical~\cite{Scott} and in real
experiments~\cite{instab-exper}.

In this Letter we show that by properly controlling the
instabilities of the system it is possible to achieve long-living
BO of matter waves in OLs. In our approach BO are implemented
using solitonic propagation (the situation similar to one explored
in truly discrete systems~\cite{Bloch-discr}). To this end we use
spatial periodic modulations of the scattering length via
optically induced Feshbach resonances, in order to change the
stability properties of the Bloch states at the edges of the band.
These modulations correspond to a nonlinear lattice whose
amplitude, considered as a free parameter, can be used to
eliminate instabilities from the band (similar stabilizing
properties of nonlinear lattice were used to achieve delocalizing
transitions in one dimensional (1D) case~\cite{BBK}). We show that
the regions of parameter space for which Bloch states become
unstable (almost) in the whole band coincide with those for which
long-lived BO of matter waves become possible, this showing  the
validity of our approach.

We start by considering the following 1D Gross-Pitaevskii equation
\begin{equation}
\label{eq:psi-in}
i\psi_t=-\psi_{xx}+\gamma x \psi+\cU(x)\psi+\cG(x)|\psi|^2\psi,
\end{equation}
describing an array of BECs in linear and nonlinear $\pi$-periodic
optical lattices of the form $\cU(x)=\cU(x+\pi)$,
$\cG(x)=\cG(x+\pi)$, respectively, with the external linear force
$\gamma$ arising from an uniform acceleration of the lattices (a
similar approach is valid also for the multidimensional case). The
consideration will be restricted to symmetric potentials
$\cU(x)=\cU(-x)$. Since BO exist in the linear regime it is
appropriate to start our considerations from the underlining
linear eigenvalue problem $-d^2\varphi_q/dx^2 +\cU(x)\varphi_q =
\varepsilon (q) \varphi_q$, where $\varphi_q$ is the standard
Bloch function corresponding to the wave vector $q$. The
periodicity of the linear OL introduces a band structure in the
spectrum e.g. the existence of chemical potential functions
$\varepsilon_n (q)$ which are periodic in reciprocal space $q\in
[-1,1]$, with $n$ denoting the band index. In the following we
shall consider only the lowest chemical potential band, a
situation typical for most BEC applications, and omit the band
index. In presence of nonlinearity, stationary localized states
(gap-solitons) can exist only if chemical potentials are inside
gaps (see e.g.~\cite{AKS}). This is not the case of a
nonstationary  (moving) soliton whose velocity, in the leading
order, coincides with the group velocity of the carrier Bloch wave
and mathematically can be described by a multiple-scale expansion
provided the linear force is weak enough $\gamma^{1/2}\ll 1$. To
this regard we search for localized solutions of
Eq.(\ref{eq:psi-in}) in the form
$\psi=\gamma^{1/2}\psi_1+\gamma\psi_2+\cdots$, where at each order
$\psi_j$ denotes a function of a set of scaled temporal
$t_n=\gamma^{n/2}t$ and spatial $x_n=\gamma^{n/2}x$ ($n=1,2,...$),
variables. Since for BO the solution, thought as a wavepacket of
Bloch states $\varphi_q$ with the carrier wave-vector depending on
the slow time $q=q(\gamma t)$, must scan the whole band, the
ansatz for the first order $\psi_1$ is chosen, in analogy with the
Houston functions~\cite{Houston} of the underlying linear theory,
of the form
\begin{eqnarray}
    \psi_1=A(\tau,\xi )e^{i\cE(t)}\varphi_{q(\tau)}(x)
\end{eqnarray}
where the phase $\cE(t)$ is determined by the equation $ d\cE/dt
=\varepsilon (q(\gamma t))$ and we have introduced specific
notations for the slow variables $\tau=\gamma t$ and
$\xi=\sqrt{\gamma} (x-v(\tau)t)$. This peculiarity of the
asymptotic expansion will manifest at the third order of the small
parameter, i.e. ${\cal O}(\gamma^{3/2})$, lower orders of the
expansion being obtained in standard manner (see e.g.~\cite{KS}).
Since the temporal dependence of the wavevector is not specified,
we impose the constraint that $q$ must follow the well known
semiclassical equation for the linear BO which in our scaled
variables acquires the simplest form $q=\tau$\footnote{This
condition allows one to cancel the third order terms originated by
the temporal dependence of the wave vector $q(\gamma t)$, and is
obtained formally by using that $\int_{0}^{\pi}\bar{\varphi}_q
\frac{\partial}{\partial q}\varphi_q
dx=\int_{0}^{\pi}x|\varphi_q|^2dx$ for symmetric potentials.}.
Then, in the order $\gamma^{3/2}$ we arrive at the NLS equation
for the slowly varying amplitude
\begin{equation}
\label{NLS}
     i  A_\tau+\frac{1}{2M(q)} A_{\xi\xi}-\chi(q) |A|^2A=0\,.
\end{equation}
Here we defined the group velocity $v(q)=d\varepsilon(q)/dq$, the
effective mass $ M(q)= (2d^2\varepsilon(q)/dq^2)^{-1} $ and the
effective nonlinearity $ \chi(q)=\int_{0}^\pi
\cG(x)|\varphi_{q(\tau)}(x)|^4dx\,. $ We emphasize that although
we have indicated $q$ as an argument in the above definitions, the
group velocity, the effective mass and the effective nonlinearity
are functions of the slow time $\tau$. Introducing "$+$" and "$-$"
subindexes to denote properties at top and bottom limits of the
band, we have that $\varphi^{(-)}=\varphi_{q=0}$, $
\varphi^{(+)}=\varphi_{q=1}$, $\chi^{(\pm)}=\int_{0}^\pi
\cG(x)|\varphi^{(\pm)}(x)|^4dx\,$, and $\mp  M^{(\pm)}>0$. As it is
clear, if $\cG(x)=$const, then $\chi^{(+)}\chi^{(-)}>0$ and from
Eq.(\ref{NLS}) it follows  that while envelope solitons are
available at one edge of the band (the edge for which $M\chi<0$),
they cannot exist at the other edge. Since for BO the soliton must
travel along the whole band, it will necessarily reach the edge
where it undergoes strong dispersion (due to defocusing action of
nonlinearity ($M\chi>0$), this leading to destruction of BO
\cite{instab-exper}.

From this analysis it is clear that an essential control parameter
of the problem is $\sigma(q)=-M(q)\chi(q)$ and in order to have
long-living BO we must require that $\sigma(q)>0$ for all $q$,
this assuring a focusing nonlinearity for the soliton dynamics
along the whole  band. The above condition can be achieved by
means of a proper design of the nonlinear lattice and will be
optimized if the condition $\sigma=$const$>0$ is satisfied.
Analytically it is easy  to consider the general case in which
$\sigma(\tau)$ is a slowly varying function $|d\sigma/dq|\ll
\sigma$. Indeed, introducing the new temporal variable
$\widetilde{\tau}=\int_{0}^{\tau} d\tau'/M(\tau')$ and the new function $\cA =
A/\sqrt{\sigma}$ we obtain the NLS equation with a dissipative
term $ i \cA_{\widetilde{\tau}}+\frac{1}{2} \cA_{\xi\xi}+
|\cA|^2\cA=i\frac{2}{\sigma}\frac{d\sigma}{dq}\cA $ whose
approximate solution can be found by means of soliton perturbation
theory~\cite{KM}. For the particular case of a static soliton in
the frame moving with the velocity $v(\tau)$, the solution reads
$\cA = \cA_0 \exp\left(\Gamma(\widetilde{\tau})+\frac i2
\cA_0^2\int_0^{\widetilde{\tau}}\sqrt{\sigma(\tau)}d\tau \right) \mbox{sech}
\left(\cA_0[\sigma(\widetilde{\tau})]^{1/4} \xi\right)$
with $\cA_0$ the constant amplitude of the soliton.

\begin{figure}
\centerline{\includegraphics[width=8.2cm,height=5.0cm,angle=0,clip]{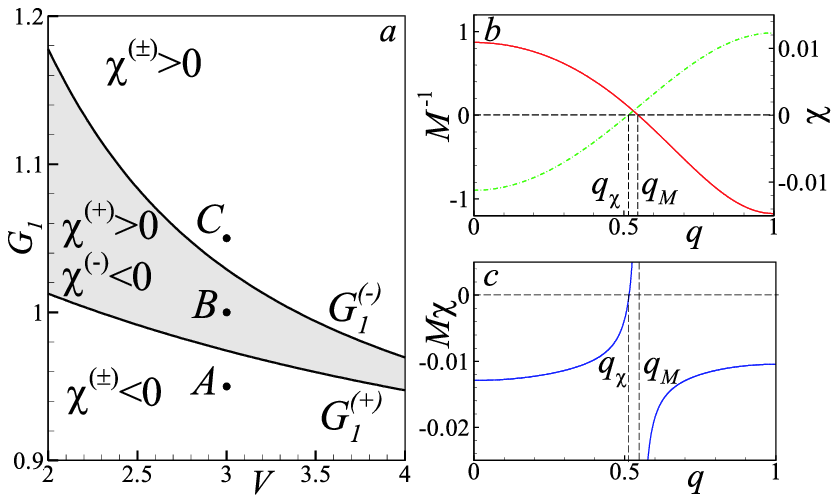}}
\caption{(a) $G_1^{(\pm)}$ {\it vs} $V$ which separate domains
where $\chi^{(\pm)}(V)\gtrless 0$.
The points A, B, and C correspond to the parameters explored below
in  Fig.\ref{fig:psi2low-X-T}b (point A),
Fig.\ref{fig:psi2low-X-T}a and \ref{fig:psi2up-X-T}a (point B),
and Fig.\ref{fig:psi2up-X-T}b (point C). Long living BO can be
observed in the shadowed domain. (b) The inverse effective mass
$M^{-1}$ (solid line) and the effective nonlinearity $\chi$
(dashed-dotted line) {\it vs} $q$.  (c)   $M\chi$ {\it vs} $q$ for
the parameters corresponding to the point B (panel a).
$q_{\chi}\approx 0.5145$ and $q_M\approx 0.5475$ indicate the
values of $q$, where $\chi$ and $M^{-1}$ are respectively equal to
zero.} \label{fig:G-V}
\end{figure}

In order to check the above predictions we have performed direct
numerical simulations of Eq. (\ref{eq:psi-in}) with
$\cU(x)=-V\cos(2x)$ and $\cG(x)=G_0+G_1\cos(2x)$, where $V$ and
$G_1$ denote the amplitudes of the linear and nonlinear lattices,
respectively, $G_0$ is the "average" nonlinearity, which in all
numerical simulations reported below is chosen to be $G_0=-0.777$.
We denote by $G_1^{(\pm)}$  the value of $G_1$ at which effective
nonlinearity $\chi^{(\pm)}$ becomes zero. Respectively if
$G_1\gtrless G_1^{(\pm)}$ we have that $\chi^{(\pm)}\gtrless 0$
and the condition for the existence of envelope solitons
(alternatively, the instability of the Bloch waves) at the both
band edges $M^{(\pm)}\chi^{(\pm)}<0$ is met when $G_1^{(+)}<
G_1<G_1^{(-)}$. This corresponds to the domain of parameters
between the lines in Fig.\ref{fig:G-V} where $G_1^{(\pm)}$ {\it
vs} $V$ are depicted (e.g. point B in Fig.\ref{fig:G-V}).
Consequently, domains of parameters $G_1<G_1^{(+)}$ (e.g. point A
in Fig.\ref{fig:G-V}) and $G_1>G_1^{(-)}$ (e.g. point C in
Fig.\ref{fig:G-V}) allow for the existence of small amplitude
envelope solitons only at lower or upper band edge,
correspondingly. Thus long-lived BOs can be expected in the domain
of parameters between the two curves $G_1^{(\pm)}(V)$ in
Fig.\ref{fig:G-V}.

Below we concentrate on the choice of the parameters corresponding
to the specific points A, B, and C, in Fig.~\ref{fig:G-V}a. These
are situations deviating form the optimally designed lattices, as
is clearly seen in panels (b) and (c) of Fig.\ref{fig:G-V}.
Indeed, while the points $q_M$
and $q_\chi$ are relatively close to each other, they do
not exactly coincide, what results in the singularity of
$\sigma(q)$ at the point $q_M$ and in the existence of the interval
$q_\chi<q<q_M$ where envelope solitons do not exist (however
outside this interval $\sigma$ is a relatively slow function of
the wavevector).

\begin{figure}
\centerline{\includegraphics[width=8.2cm,height=8.2cm,angle=0,clip]{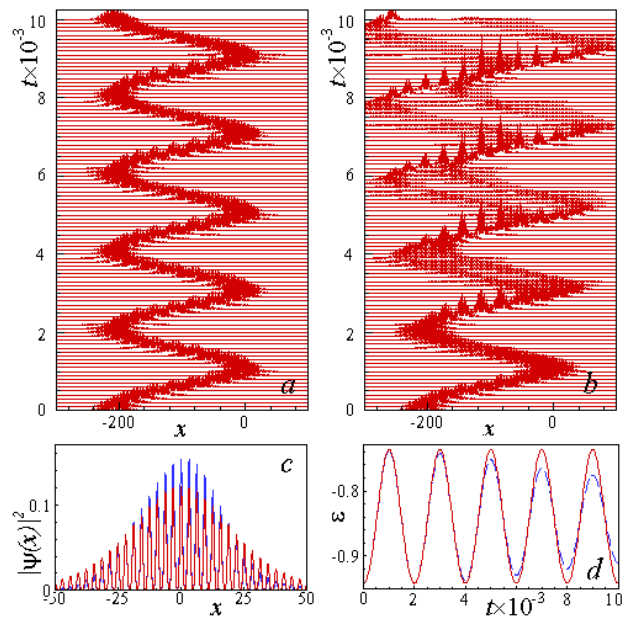}}
\caption{(Color online) (a)  Long-living BO of the envelope
soliton for $G_1=1.0$ (corresponding to point B in
Fig.\ref{fig:G-V}a). (b) Decaying BO for $G_1=0.95$ (corresponding
to point A in Fig.\ref{fig:G-V}a). In both cases oscillations
start near the bottom of the band at $\varepsilon=-0.938$ with
parameter values $\gamma=-0.001$, $V=3.0$ and with initial
condition $X_0=-65\pi$. (c) The shape of the envelope soliton in
panel (a) at time $t=9000$ (thin solid lines) is compared with the
stationary state (thick solid lines) with the same number of
particles $N=2.42$ and chemical potential nearby the upper edge of
the band at $\varepsilon=-0.7324$. (d) The chemical potential
$\varepsilon(t)$ for long-living ($G_1=1.0$, solid line) and
decaying ($G_1=0.95$, dashed line) BO. The lowest allowed band
corresponds to the interval $\varepsilon\in[-0.93683,-0.73326]$. }
\label{fig:psi2low-X-T}
\end{figure}

In Fig.\ref{fig:psi2low-X-T}a we present the time evolution of a
small amplitude envelope soliton obtained by direct numerical
integration of Eq.(\ref{eq:psi-in}). As initial condition we used
the stationary envelope soliton of Eq. (\ref{eq:psi-in}) with
$\gamma=0$, whose chemical potential belongs to the semi-infinite
gap very close to the lowest allowed band. From this figure the
existence of  long-living BO with the period $T_s \approx 2\cdot
10^3$ and spatial amplitude $X_s \approx 32.5\pi$, perfectly
matching the semiclassical  estimates $2/|\gamma|$ and
$(\varepsilon^{(+)}-\varepsilon^{(-)})/(2|\gamma|)$, respectively,
is quite evident. Notice that the turning points of the spatial
dynamics correspond to values of chemical potentials inside the
gaps (very close to band edges) for which stationary solitons
exist. In Fig.\ref{fig:psi2low-X-T}c we have compared  the shape
of the soliton at the fifth right turning points of the BO with
the profile of the stationary state at the top of the band, from
which we see that they  practically coincide (the small difference
is ascribed to the radiative effects of the BO dynamics).
Remarkably accurate prediction of the theory is verified by
comparing the dynamical profiles of the soliton at the turning
points of the BO. The profiles at the times $t=10^3;\,3\cdot
10^3;\,5\cdot 10^3;\, 7\cdot 10^3$ were practically
indistinguishable from the one depicted for $t=9\cdot 10^3$ in
Fig.\ref{fig:psi2low-X-T}c. Numerical simulations performed on
longer time scales (up to $t=2\cdot 10^4$) showed no appreciable
decay of the BO and perfect recovering of the soliton shape at the
turning points.

\begin{figure}
\centerline{\includegraphics[width=8.2cm,height=8.2cm,angle=0,clip]{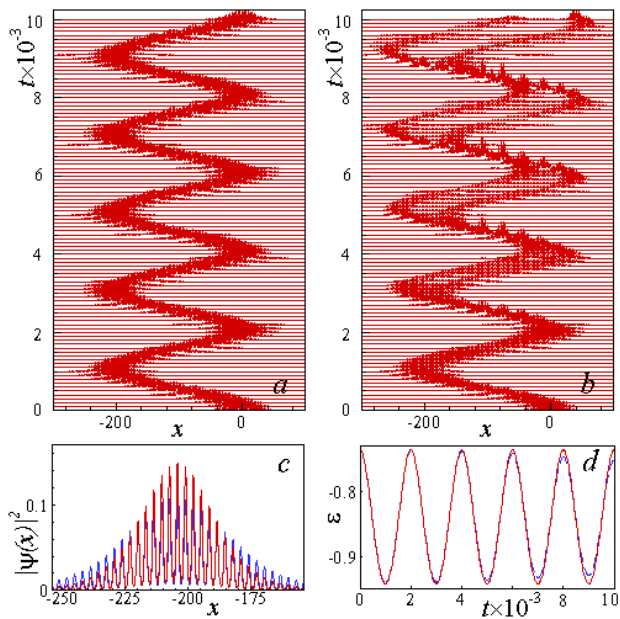}}
\caption{ (Color online)  Dynamics of BO for the cases  (a)
$G_1=1.0$ and (b) $G_1=1.05$ (corresponding, respectively, to
points B and A in Fig.\ref{fig:G-V}a). In both cases the
oscillations start near the top of the band at
$\varepsilon=-0.7324$, with $\gamma=-0.001$, $V=3.0$ and initial
condition $X_0=0\pi$. In (c) we compare the profile of the soliton
at $t=9\cdot 10^3$ in panel (a) (thin solid lines) with the
stationary solution (thick solid lines) with the same number of
particles $N=2.42$ and belonging to the semi-infinite gap
($\varepsilon=-0.938$). In (d) the dynamics of the chemical
potential $\varepsilon(t)$ is shown for $G_1=1.0$ (solid line) and
$G_1=1.05$ (dashed line). } \label{fig:psi2up-X-T}
\end{figure}

To appreciate the importance of the parameter design implied by
our theory, we depict in Fig.\ref{fig:psi2low-X-T}b the dynamics
of a soliton corresponding to point A in Fig.\ref{fig:G-V}. We see
that in this case the BO undergo  fast decay due do spreading  of
the wave packet. The difference in the two types of BO is clearly
seen also from the dependence of the chemical potential on time,
computed as $\varepsilon(t)=\frac 1 N
\int_{-\infty}^{\infty}\left(\left|
\psi_{x}\right|^2+\cU(x)\left|\psi \right|^2+\cG(x)\left|\psi
\right|^4\right)dx $, with
$N=\int_{-\infty}^{\infty}\left|\psi\right|^2 dx$ denoting the
number of atoms. This is shown in Fig.\ref{fig:psi2low-X-T}d from
which we see that stable BO correspond to perfectly periodic
trajectory (solid line) while decaying oscillations correspond to
decays of the oscillations of the chemical potential (dashed
line).
Similar phenomena exist also if the BO is started from stationary
states at the top, rather than at the bottom, of the band for the
same  parameter values of points $B,C$, in Fig.\ref{fig:G-V}a), as
one can see from Fig.\ref{fig:psi2up-X-T}.

In closing this letter we wish to discuss possible experimental
settings for observing long-living BO in BECs trapped in uniformly
accelerated linear and nonlinear OLs  (with acceleration
$2\gamma$): $\cU(X-\gamma T^2)$ and $\cG(X-\gamma T^2)$,
respectively. The corresponding model equation in dimensionless
variables takes the form
\begin{eqnarray}
\label{eq:psi-acc} i\Psi_T=-\Psi_{XX}+\cU(X-\gamma
T^2)\Psi+\cG(X-\gamma T^2)|\Psi|^2\Psi.
\end{eqnarray}
After substituting $\psi=e^{-i\left[(X-\gamma T^2)\gamma
T+\gamma^2T^3/3\right]}\Psi$ and introducing new independent
variables $x=X-\gamma T^2$ and $t=T$, Eq. (\ref{eq:psi-acc}) takes
the form of a NLS equation with external linear force
(\ref{eq:psi-in}). In this normalization the energy is measured in
units of the recoil energy $E_r=\hbar^2\pi^2/(2m d^2)$, where $d$
is the lattice period and $m$ is the mass of bosons, and the
spatial and temporal variables are measured in units of $d/\pi$
and $\hbar/E_r$, respectively. To check the experimental
feasibility of the proposed setting, we consider a $^{87}{\rm Rb}$
condensate in a trap with a transverse radial size $a=2\,\mu {\rm
m}$ and with a period of linear and nonlinear lattices $d=1\,\mu
{\rm m}$. Then dimensionless parameters of long-living BO,
depicted in Fig.\ref{fig:psi2low-X-T}a and \ref{fig:psi2up-X-T}a
will correspond to soliton, containing $N\approx 3800$  atoms,
driven by external force $1.17\cdot 10^{-27} {\rm N}$ (which is
obtained, when the acceleration of linear and nonlinear lattices
is of order of $\sim 8 {\rm mm/s}$). At the same time the above
parameters of the nonlinear lattice could be created by the
spatial variation (obtained by the optically induced Feshbach
resonance) of the bosonic s-wave scattering length
$a_s(x)=a_s^{(0)}+a_s^{(1)}\cos{(2\pi x/d)}$ with the "average"
value $a_s^{(0)}=-1.554 \,{\rm nm}$ and the amplitude $a_s^{(1)}=2
\,{\rm nm}$. For the parameters of Fig.\ref{fig:G-V}, the
long-living BO will occur when the amplitude of the spatial
variation of the scattering length is in the range $1.948 \,{\rm
nm} < a_s^{(1)} < 2.058 \,{\rm nm}$. This show that the long lived
BO reported in this Letter can indeed  be observed in the
experimental settings available today.

\smallskip
M.S. acknowledges
support from a MUR-PRIN-2005 initiative {\it Transport properties
of classical and quantum systems}.  VVK was supported by the
FCT and European program FEDER under Grant No.
POCI/FIS/56237/2004. Y.V.B. acknowledges support of
the FCT under the Grant No. SFRH/PD/20292/2004.

\end{document}